\documentstyle[psfig]{mn}

\newcommand\mdot   {\hbox {${\dot M}$}}

\newcommand\mzon   {M$_{\odot}$}
\newcommand\pp     {$\pm$}

\newcommand\micros  {$\mu$s}

\title[The 5 Hz QPO in GRS 1739--278]{A detailed study of the 5 Hz
quasi-periodic oscillations in the bright X-ray transient and
black-hole candidate GRS 1739--278}

\author[Wijnands et al.]{Rudy Wijnands$^1$\thanks{Chandra Fellow;
       Email: rudy@space.mit.edu},  Mariano
       M\'endez$^2$, Jon M. Miller $^1$, Jeroen Homan$^3$\\ 
       $^1$ Center for Space Research, Massachusetts Institute of
       Technology, 77 Massachusetts Avenue, Cambridge, MA 02139-4307,
       USA\\
       $^2$ SRON Laboratory for Space Research, Sorbonnelaan 2,
       NL-3584 CA, Utrecht, The Netherlands\\
       $^3$ Astronomical Institute ``Anton Pannekoek'', University of
       Amsterdam, Kruislaan 403, NL-1098 SJ Amsterdam, The Netherlands}

\begin{document}
\maketitle

\begin{abstract}
We present a detailed study of the 5 Hz quasi-periodic oscillation
(QPO) recently discovered in the bright X-ray transient and black-hole
candidate GRS 1739--278 (Borozdin \& Trudolyubov 2000) during a {\it
Rossi X-ray Timing Explorer} observation taken on 1996 March 31.  In
total 6.6 ksec of on-source data were obtained, divided in two data
sets of 3.4 and 3.2 ksec which were separated by $\sim$2.6 ksec. The 5
Hz QPO was only present during the second data set. The QPO increased
in strength from below 2\% rms amplitude for photon energies below 4
keV to $\sim$5\% rms amplitude for energies above 10 keV. The soft QPO
photons (below 5 keV) lagged the hard ones (above 10 keV) by almost
1.5 radian. Besides the QPO fundamental, its first overtone was
detected. The strength of the overtone increased with photon energy
(from $<$2\% rms below 5 keV to $\sim$8\% rms above 10 keV). Although
the limited statistics did not allow for an accurate determination of
the lags of the first overtone, indications are that also for this QPO
the soft photons lagged the hard ones. When the 5 Hz QPO was not
detected (i.e., during the first part of the observation), a broad
noise component was found for photon energies below 10 keV but it
became almost a true QPO (with a Q value of $\sim$1.9) above that
energy, with a frequency of $\sim$ 3 Hz. Its hard photons preceded the
soft ones in a way reminiscent of the 5 Hz QPO, strongly suggesting
that both features are physically related. We discuss our finding in
the frame work of low-frequency QPOs and their properties in BHCs.

\end{abstract}


\section{Introduction \label{intro}}

Before the launch of the {\it Rossi X-ray Timing Explorer} ({\it
RXTE}), the usual picture for black-hole candidates (BHCs) was simple
and one-dimensional: the changes in the X-ray spectra and the rapid
X-ray variability are caused by changes in the mass accretion rate
\mdot~(Tanaka \& Lewin 1995; van der Klis 1995). In the BHC low-state
(LS), \mdot~is low and the spectra are hard. The LS power spectra are
dominated by a very strong (20\%--50\% rms amplitude) band-limited
noise which follows approximately a power-law with index 1 at high
frequencies, but below a certain frequency (the break frequency) the
power spectrum becomes roughly flat. Above this frequency a broad bump
or a quasi-periodic oscillation (QPO) is often present, although also
QPOs with similar frequencies as the break frequency are observed.  In
some sources, a second break is visible in the power spectrum, above
which the power-law index increases to about $\sim$2. In the high
state (HS), \mdot~is higher, the spectra are much softer, and in the
power spectra only a weak (a few percent) power-law noise component is
present. In the very high state (VHS), \mdot~is the highest, the
spectra are harder but not as hard as in the LS, and in the power
spectra, noise is present similar to either the weak HS noise or the
LS band-limited noise (although only with a strength of 1\%--15\%
rms).  QPOs near 6 Hz are detected sometimes with a complex harmonic
structure.

With the many observations performed with {\it RXTE} on BHCs, the
behaviour of BHCs turned out to be much more complex than previously
thought. First of all, {\it RXTE} has expanded the range of
frequencies at which the BHCs show variability up to 450 Hz (e.g.,
Remillard et al. 1999a,b; Cui et al. 2000; Homan et al. 2001;
Strohmayer 2001). The nature of these BHC high-frequency QPOs is
unclear, although it has been suggested that they might be related to
the twin kHz QPOs in the neutron star systems (Psaltis et
al. 1999). Also, with {\it RXTE} in many BHCs the VHS has now been
detected and the associate low-frequency ($<$20 Hz) QPOs have been
found (e.g., Remillard et al. 1999b; Borozdin \& Trudolyubov 2000;
Revnivtsev, Trudolyubov, \& Borozdin 2000; Dieters et al. 2000; Cui et
al. 2000; Homan et al. 2001), indicating that these QPOs are a common
feature of BHCs. The phenomenology of these QPOs is very complex and
they are observed during different luminosity levels (at levels
significantly below the highest observed luminosities, i.e., not only
during the VHS but at states intermediate between the LS and the HS;
e.g., Homan et al. 2001).  The phenomenology of these QPOs is such
that even in individual sources, the QPO interrelationships are not
fully understood (see, e.g., Wijnands et al. 1999 or Homan et
al. 2001).  The relationship between the QPOs observed in different
BHCs is even less well understood.  However, from a detailed study of
the state behaviour in XTE J1550--564 it is clear that the
one-dimensional picture described above for the BHC states (depending
only on \mdot) does not hold in this particular source and another
extra parameter is needed to explain its behaviour (Homan et
al. 2001). Similar behaviour might also be observable for the other
BHCs, however, not much information about this is presently available.

One of the most recent additions to the collection of BHCs which
exhibit QPOs near 6 Hz is the X-ray transient and BHC GRS 1739--278.
At the end of March 1996, this new X-ray transient was discovered with
the SIGMA telescope on board {\it Granat} (Paul et al. 1996; Vargas et
al. 1997). Soon after the discovery of this new X-ray source its radio
and optical counterpart where discovered (Hjellming et al. 1996;
Mirabel et al. 1996; Marti et al. 1997). From a {\it ROSAT}
observation of the source an extinction of $A_{\rm v}= 14$\pp2 was
derived, implying that the source is located at a distance of 6--8.5
kpc and that it was radiating at least near the Eddington limit for a
1 \mzon~compact object (Dennerl \& Greiner 1996; Greiner, Dennerl, \&
Predehl 1996). In X-rays, GRS 1739--278 was studied by the TTM
experiment on board {\it Mir-Kvant} and by the PCA instrument on board
{\it RXTE} (Borozdin, Alexandrovich, \& Sunyaev 1996; Takeshima et
al. 1996; Borozdin et al. 1998). The {\it RXTE} data showed that on
1996 March 31, the source was in the canonical black-hole very high
state and the {\it RXTE} observations taken in 1996 May showed that
the source had transit to the canonical high state. Recently, Borozdin
\& Trudolyubov (2000) reported the discovery of a 5 Hz QPO during the
1996 March 31 {\it RXTE} observation. Because the QPO behaviour of
BHCs is not well understood, we decided to study the properties of
this 5 Hz QPO in GRS 1739--278 in more detail (e.g., time variability,
phase lags).

\section{Observation and analysis \label{observations}}

\begin{figure}
\begin{center}
\begin{tabular}{c}
\psfig{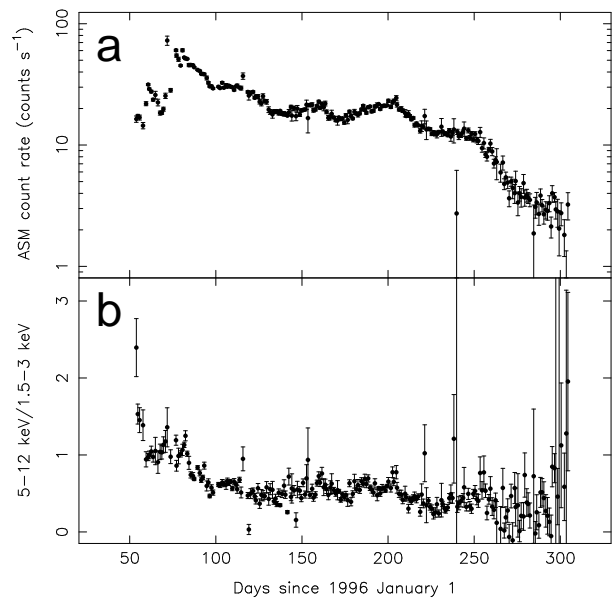}
\end{tabular}
\caption{The daily-averaged {\it RXTE}/ASM light curve of GRS
1739--278 ({\it a}) and the 5--12 keV/1.5--3 keV count rate ratio
curve of the source ({\it b}).  The 5 Hz QPO was detected on 1996
March 31 which is day 90 in this figure.
\label{fig:asm} }
\end{center}
\end{figure}

{\it RXTE} observed GRS 1739--278 on several occasions during its 1996
outburst.  For a detailed overview of the {\it RXTE} observations, we
refer to Borozdin et al. (1998). In this paper, we concentrate on the
1996 March 31 observation during which the 5 Hz QPO was discovered
(Borozdin \& Trudolyubov 2000). In order to avoid confusion with the
bright neutron-star X-ray transient GRO J1744--28, the satellite was
not pointed directly at GRS 1739--278 but had an off-set pointing of
$\sim11$ arcmin (see Takeshima et al. 1996, Borozdin et
al. 1998). Borozdin \& Trudolyubov (2000) showed that the QPO almost
certainly originated from GRS 1739--278 and not from GRO J1744--28.

During the 1996 March 31 observation, data were simultaneously
collected in the Standard2f mode (129 photon energy channels with 16 s
time resolution), in the single bit modes SB\_125US\_0\_13\_1S and
SB\_125US\_14\_35\_1S (1 photon energy channel, 128 \micros~time
resolution), in the binned mode B\_2MS\_16A\_0\_35\_Q (16 photon
energy channels, 2 ms time resolution), and in the event mode
E\_16US\_64M\_36\_1S (64 photon energy channel, 16 \micros~ time
resolution). We used the Standard2f data to create a light curve and
the colours. For this analysis, we only used the data during which all
5 detectors were on; because of this restriction, we disregarded the
first $\sim$1000 seconds of the observation during which 2 out of the
5 detectors were off.  For the analysis of the rapid X-ray variability
we used all data available, unless otherwise noted.  The high time
resolution modes (single bit, binned, and event mode data) were used
to calculate discrete Fourier transforms between 1/16--4096 Hz or
1/16--256 Hz, from which the power and cross spectra were
calculated. The power spectra were fitted with a constant
(representing the dead-time modified Poisson noise), a power-law (for
the low-frequency noise component), and one or more Lorentzians (for
the peaked noise component and the QPOs). The uncertainties in the fit
parameters were calculated using $\Delta\chi^2 = 1$ and upper limits
using $\Delta \chi^2 = 2.7$ which yields 95\% confidence levels.  The
phase lags of the QPOs were calculated for a frequency interval equal
to the FWHM of the Lorentzians used to fit the QPOs, centred on the
peak frequency of these Lorentzians.  To correct for the small
dead-time effects on the phase lags, we subtracted the average 50--125
Hz cross vector from the cross spectra (see van der Klis et
al. 1987). The single bit and the event mode data were used to search
for high frequency ($>$100 Hz) QPOs.

\section{Results}

\begin{figure}
\begin{center}
\begin{tabular}{c}
\psfig{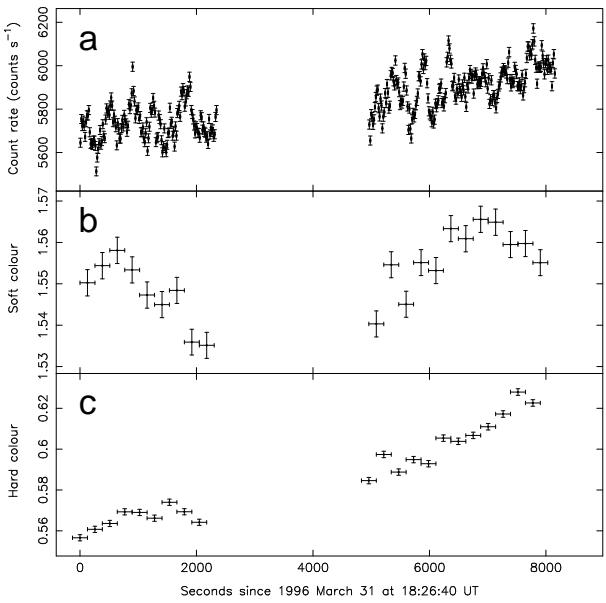}
\end{tabular}
\caption{The 2--60 keV light curve ({\it a}), the soft colour ({\it
b}), and the hard colour ({\it c}) of GRS 1739-278 on 1996 March
31. The first $\sim$1000 seconds of the observation were not used
because only 3 out of the 5 detectors were on. The soft colour is the
3.7--6.2 keV/2.2--3.7 keV count rate ratio and hard colour the
6.2--16.2 keV/2.2--3.7 keV count rate ratio. The time resolution for
the light curve is 16 seconds in order to show the rapid count rate
variations. The time resolution of the colours is 256 seconds in order
to decrease the errors on the colours. The count rates were corrected
for background and off-set (the collimator correction factor was
1.1572) but not for dead-time ($\sim$15\%). The gap in the data is due
to an Earth occultation of the source and a passages of the satellite
through the SAA.
\label{fig:lc_colours} }
\end{center}
\end{figure}

\subsection{Light curve and colours}
\begin{figure}
\begin{center}
\begin{tabular}{c}
\psfig{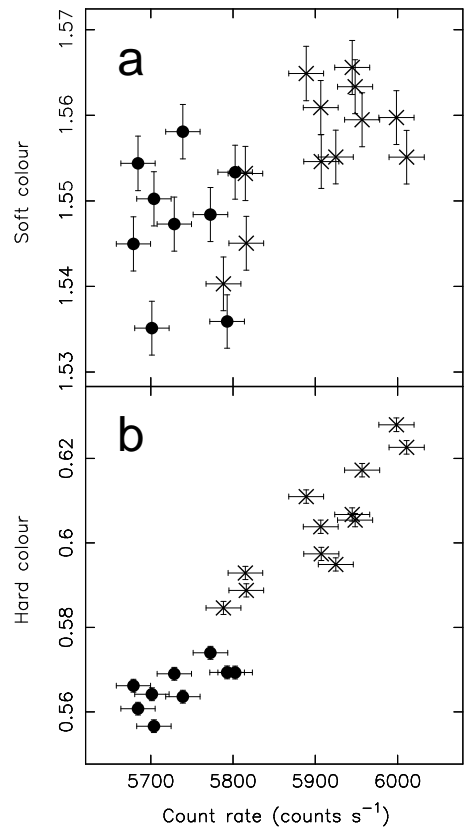}
\end{tabular}
\caption{Hardness-intensity diagrams (HIDs) of GRS 1739-278 on 1996
March 31: ({\it a}) the 'soft' HID and ({\it b}) the 'hard' HID.  The
first $\sim$1000 seconds of the observation were not used because only
3 out of the 5 detectors were on. The soft colour is the 3.7--6.2
keV/2.2--3.7 keV count rate ratio and hard colour the 6.2--16.2
keV/2.2--3.7 keV count rate ratio.  The count rate was measured in the
2--60 keV energy range and was corrected for the satellite off-set of
$\sim$11 arcmin. The time resolution is 256 seconds. The count rates
were corrected for background but not for dead-time. The solid dots
are the data for the first part of the observation; the crosses for
the second part.\label{fig:hid} }
\end{center}
\end{figure}

The {\it RXTE} all sky monitor (ASM\footnote{The quick-look ASM data
used in this paper can be obtained from
http://xte.mit.edu/ASM\_lc.html and are provided by the ASM/{\it RXTE}
team. See Levine et al. (1996) for a detailed description of the
ASM.}) light curve (1.5--12 keV) of the source is shown in
Figure~\ref{fig:asm}{\it a} (see also Borozdin et al. 1998; Borozdin
\& Trudolyubov 2000). The outburst light curve of GRS 1739--278 can be
classified as a typical fast rise, exponential decay light curve,
although with multiple maxima in the decay. At the peak of the
outburst the flux was about 1 Crab, which makes this source a bright
X-ray transient. The 5--12 keV/1.5--3 keV count rate ratio curve is
shown in Figure~\ref{fig:asm}{\it b}, which shows that the spectrum
was hardest at the start of the outbursts (during the rise, before the
peak of the outburst) and gradually became softer. The 5 Hz QPO was
discovered during the 1996 March 31 {\it RXTE} observation, which is
day 90 in this figure, at a time when the source was already
considerable weaker than at the peak of the outburst, and the X-ray
spectrum had significantly softened compared to previous days.

The 2--60 keV light curve of the {\it RXTE}/PCA observation of 1996
March 31 is shown in Figure \ref{fig:lc_colours}{\it a}. The gap
between the two parts of the light curve is due to an Earth
occultation of the source and a passage of the satellite trough the
South Atlantic Anomaly (SAA). The light curve has been corrected for
the instrument off-set (the collimator correction factor was 1.1572).
In Figure \ref{fig:lc_colours}{\it b} and {\it c}, the soft and the
hard colour curves are shown, respectively. As soft colour we used the
3.7--6.2 keV/2.2--3.7 keV count rate ratio and as hard colour we used
the 6.2--16.2 keV/2.2--3.7 keV count rate ratio.

\begin{figure}
\begin{center}
\begin{tabular}{c}
\psfig{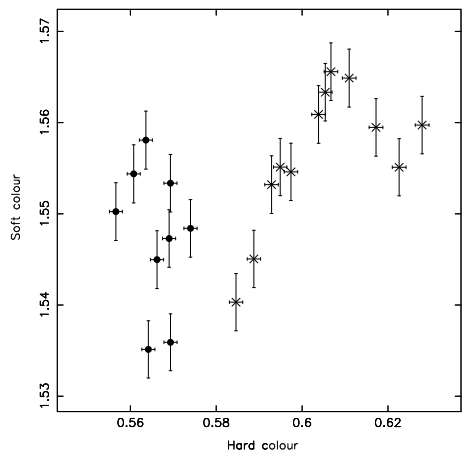}
\end{tabular}
\caption{Colour-colour diagram of GRS 1739-278 on 1996 March 31. The
first $\sim$1000 seconds of the observation were not used because only
3 out of the 5 detectors were on. The soft colour is the 3.7--6.2
keV/2.2--3.7 keV count rate ratio and hard colour the 6.2--16.2
keV/2.2--3.7 keV count rate ratio.  The time resolution is 256
seconds. The data were corrected for back-ground.  The solid dots
correspond to the first part of the observation; the crosses to the
second part.
\label{fig:cd} }
\end{center}
\end{figure}

From Figure~\ref{fig:lc_colours}{\it a}, it is apparent that the
source was highly variable on time scales of minutes. Also, it can be
observed that, although the source count rate was gradually decreasing
on time scales of weeks, the count rate increased with time during
this observation, which is most clearly visible in the second part of
the observation. The soft colour (Fig. \ref{fig:lc_colours}{\it b})
did not show a clear correlation with time, whereas the hard colour
(Fig. \ref{fig:lc_colours}{\it c}) clearly increased with time (best
visible in the second part of the observation). Because the count rate
also increased with time (Fig.~\ref{fig:lc_colours}{\it a}), in the
'hard' hardness-intensity diagram (HID; as colour the hard colour was
used; Fig. \ref{fig:hid}{\it b}), the hard colour showed a clear
positive correlation with the count rate, but in the 'soft' HID (as
colour the soft colour was used; Fig. \ref{fig:hid}{\it a}), the soft
colour did not show such a good correlation (as expected from the lack
of a good correlation of the soft colour with time), although also the
soft colour seemed to slightly increase with count rate. This is
reflected in the colour-colour diagram (CD; Fig. \ref{fig:cd}) which
shows that the soft colour tended to increase when also the hard
colour increased.

\subsection{Rapid X-ray variability}

\begin{figure}
\begin{center}
\begin{tabular}{c}
\psfig{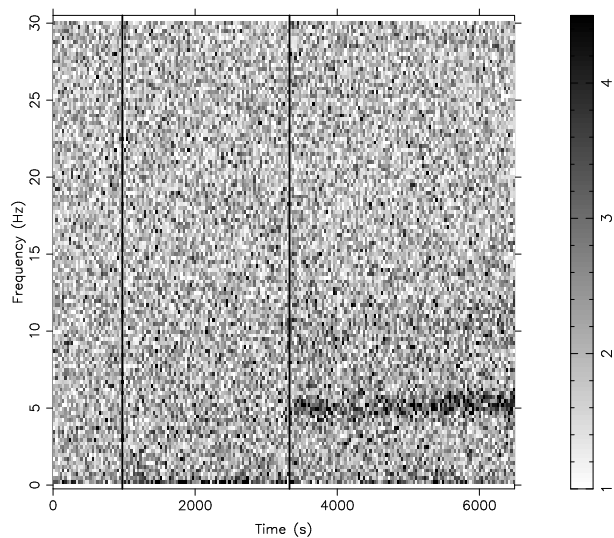}
\end{tabular}
\caption{Dynamical power spectrum of the whole observation, including
the part were only 3 detectors were on. The time resolution of the
data is 32 seconds. The first black line indicates when the number of
detectors that were on changed from 3 to 5. The second black line
indicates the gap between the two data parts due to Earth occultation
of the source and a passage of the satellite trough the SAA
($\sim$2600 seconds).
\label{fig:dynamical} }
\end{center}
\end{figure}

\begin{figure}
\begin{center}
\begin{tabular}{c}
\psfig{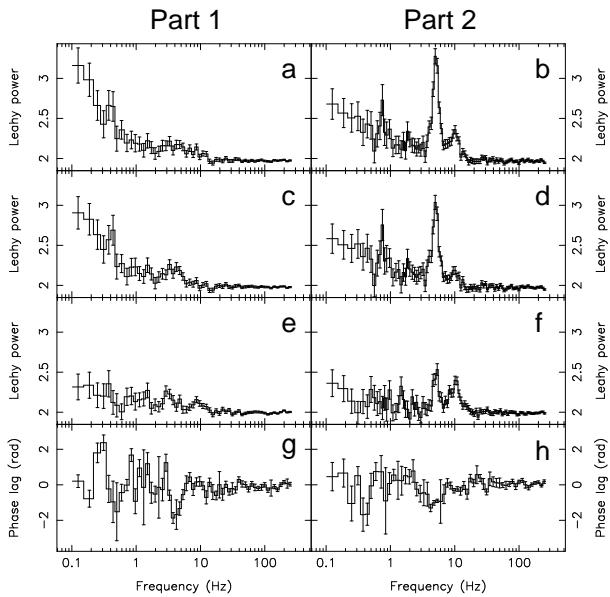}
\end{tabular}
\caption{Leahy normalized (Leahy et al. 1983) power spectra of the
first ({\it a}, {\it c}, and {\it e}) and second ({\it b}, {\it d},
and {\it e}) part of the observation in the energy range of 2--60,
2--8.7, and 8.7--60 keV, respectively. The Poisson level has been not
been subtracted.  Cross spectra of the first ({\it g}) and second
({\it h}) part of the observation, calculated between 2.8--8.7 and
8.7--60 keV. Negative signs mean that the hard photons precede the
soft ones.
\label{fig:powerspectra}}
\end{center}
\end{figure}

In order to study the rapid X-ray variability (i.e., the 5 Hz QPO
discovered by Borozdin \& Trudolyubov 2000), we first produced a
dynamical power spectrum of the data (Fig. \ref{fig:dynamical}). From
this figure, it is apparent that the QPO was only clearly present in
the second part of the data and not in the first part. To check this,
we created a power spectrum for each part separately (shown in
Figures~\ref{fig:powerspectra}{\it a} and {\it b}).  Using the energy
range 2.8--31.7 keV, a very significant 5 Hz QPO (rms amplitude
2.33\%\pp0.07\% [$\sim18\sigma$]; frequency 5.09\pp0.03 Hz, width
1.11\pp0.09 Hz) and its first overtone (rms amplitude 1.9\pp0.1
[$\sim9\sigma$]; frequency 10.2\pp0.2 Hz; width 2.7\pp0.4 Hz) were
present during the second part of the observation, but during the
first part the 5 Hz QPO was not present (but see below), although a
weak noise component around a few hertz was (see
Fig.~\ref{fig:powerspectra}). During the second part, an extra noise
component at low frequencies was also present which was fitted with a
power-law (rms amplitude 1.16\%\pp0.09\% and with a power-law index of
0.71\pp0.08).

\begin{figure}
\begin{center}
\begin{tabular}{c}
\psfig{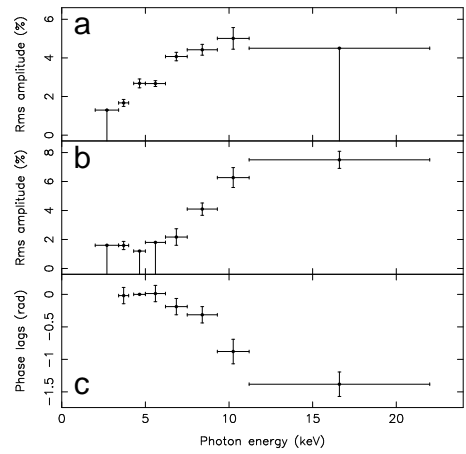}
\end{tabular}
\caption{The rms amplitude of the fundamental ({\it a}), the rms
amplitude of the overtone ({\it b}), and the phase lags of the
fundamental ({\it c}) versus photon energy. In ({\it c}) the reference
band chosen was 4.3--5.0 keV. Negative signs mean that the hard
photons precede the soft ones. \label{fig:energy_lags}}
\end{center}
\end{figure}

To study the energy dependence of the rapid X-ray variability, we made
power spectra of each part in a soft band (2--8.7 keV;
Figs. \ref{fig:powerspectra}{\it c} and {\it d}) and a hard one
(8.7--60 keV; Figs. \ref{fig:powerspectra}{\it e} and {\it f}). At
first glance, the noise component in the first part of the observation
did not show any clear dependence on energy (but see below), but the
QPOs in the second part did: at low energies the fundamental was
stronger than its overtone, but at higher energies they were of
comparable strength.  The exact dependence of the QPO amplitude on
photon energy is shown in Figure~\ref{fig:energy_lags}{\it a} for the
fundamental and in Figure \ref{fig:energy_lags}{\it b} for the first
overtone (see also Tab.~\ref{tab:energy}). From these figures, it is
clear that at the highest energies ($>10$ keV) the overtone was even
stronger than the fundamental, demonstrating that the overtone had a
steeper dependence on photon energy than the fundamental.

\begin{figure}
\begin{center}
\begin{tabular}{c}
\psfig{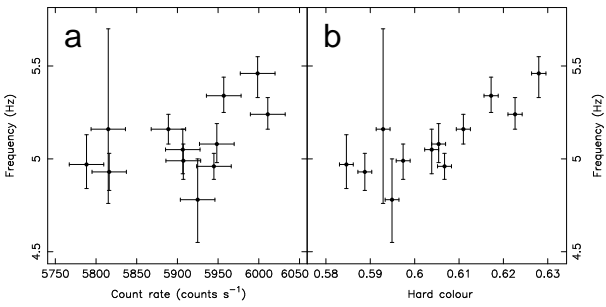}
\end{tabular}
\caption{The frequency of the fundamental versus the 2--60 keV count
rate ({\it a}) and the hard colour ({\it b}). The count rates were
corrected for back-ground and collimator off-set, but not for
dead-time. The hard colour is the 6.2--16.2 keV/2.2--3.7 keV count
rate ratio.
\label{fig:qpo_lc_hc}}
\end{center}
\end{figure}

To study the behaviour of the QPO fundamental in the second part with
count rate and colours, we used 256 s data segments to track the QPO
parameters in time. The results are shown in
Figure~\ref{fig:qpo_lc_hc}.  The frequency of the QPO increased
slightly when the count rate increased (Fig. \ref{fig:qpo_lc_hc}{\it
a}) but more strongly when the hard colour increased
(Fig. \ref{fig:qpo_lc_hc}{\it b}), suggesting that the hard colour is
a better tracer of the QPO frequency than the count rate. The other
parameters of the QPO (rms amplitude and FWHM) did not show clear
correlations either with the count rate or the colours.

We calculated cross spectra in order to study the phase lags of the
QPOs. The phase lags between two broad energy bands as a function of
frequency is shown in Figures~\ref{fig:powerspectra}{\it g} and {\it
h}. Clearly the 5 Hz QPO had negative phase lags
(Fig. \ref{fig:powerspectra}{\it h}), meaning that the hard photons of
the 5 Hz QPO preceded the soft ones. The phase lags of the 5 Hz QPO as
a function of photon energy are shown in
Figure~\ref{fig:energy_lags}{\it c}. The lags increased with
increasing photon energy. The lowest energy bin is not shown because
the QPO was not detected in this energy band.  Using two broad energy
bands, we find a $\sim$3$\sigma$ significant phase lag of
--0.26\pp0.08 rad. for the first overtone between the energy bands
2.8--8.7 and 8.7--60 keV. Although barely significant, this might
indicate that also for the first overtone the hard photons preceded
the soft ones.

\begin{figure}
\begin{center}
\begin{tabular}{c}
\psfig{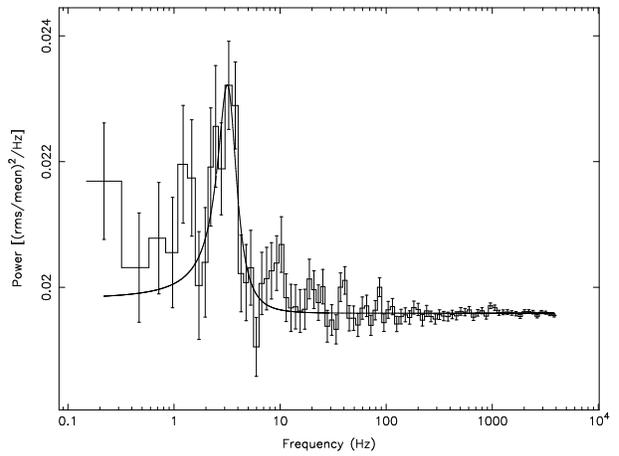}
\end{tabular}
\caption{Power spectrum for the energy range 11.2--31.7 keV of the
first data set (only the data for which all 5 PCUs were on) showing
the 3 Hz QPO. The Poisson level has not been subtracted.
\label{fig:3Hzqpo}}
\end{center}
\end{figure}

Figure \ref{fig:powerspectra}{\it g} shows that the noise component
around 3--5 Hz in the first part of the data also had negative phase
lags. The profile in phase lag diagram is rather narrow (narrower than
the width of the noise component) and suggest that it might be due to
a narrower noise component or even a QPO. To investigate this
possibility, we examined the power spectra of the first part of the
data in great detail for different energy ranges.  We found that the
noise component becomes more peaked at higher energies until it was
nearly a true QPO (with a Q value of $\sim$1.9; hereafter, we refer to
this feature as QPO). In Figure~\ref{fig:3Hzqpo}, the power spectrum
of the first part of the data is shown (only when all 5 PCUs were on)
for the energy range 11.2--31.7 keV containing the QPO. Its strength
was 9.8$^{+0.8}_{-0.7}$\% rms amplitude (6.7$\sigma$ significance),
its FWHM was 1.7$^{+0.5}_{-0.3}$ Hz, and its frequency was 3.2\pp0.1
Hz. Below 11.2 keV either this QPO was significantly broader or an
extra noise component was present, which makes the QPO upper limits at
lower energies difficult to obtain. A conservative upper limit can be
given by assuming that all the power present is due to the QPO. This
results in an upper limit of 4.5\% and 1.1\% for the energy ranges
4.3--11.2 keV and 2--4.3 keV, respectively. The QPO phase lag between
2.8--8.7 and 8.7--60 keV was --2.4\pp0.2 radian, which absolute value
is larger than the absolute value of the phase lag of the 5 Hz QPO
(see Tab.~\ref{tab:diff}). For both QPOs the hard photons precede the
soft ones.

Recently, QPOs above 100 Hz were found in several BHC, simultaneously
with QPOs round 5--7 Hz (e.g., Homan et al. 2001; Cui et
al. 2000). We searched the data for similar QPOs in GRS 1739--278, but
none were found. The 95\% confidence upper limits between 100 and 1000
Hz are 2.3\% rms (1.7\% rms) and 2.3\% rms (1.8\% rms), respectively
for 2--60 keV and 4.3--60 keV, assuming a width of 150 Hz (the values
for a FWHM of 50 Hz are given between brackets).

\section{Discussion \label{discussion}}

We present a detailed study of the 5 Hz QPO in the bright X-ray
transient and black-hole candidate GRS 1739--278. This QPO was only
observed during one {\it RXTE} observation (on 1996 March 31) during
which the source was in state which can be identified as a very-high
state (see also Borozdin et al. 1998). However, it is clear that the
source was not at its highest possible luminosity demonstrating that
very-high state behaviour can also occur at low luminosities, similar
to what has been found in other BHCs (e.g., Homan et al. 2001).  We
showed that only during the second part of the observation the 5 Hz
QPO and its first overtone were prominently present. During the first
part only a broad noise component was present for energies below 10
keV and a broad QPO near 3 Hz for energies above that.  All the QPOs
considerably increased in strength with increasing photon energies,
and the phase lags for all the QPOs demonstrate that the hard photons
preceded the soft ones by as much as 1.5--2.5 radian.

The similarities between the 3 Hz QPO and the 5 Hz QPO (see
Tab.~\ref{tab:diff}) strongly suggests that they are directly related
to each other. Most likely, the 3 Hz QPO evolved in the 5 Hz QPO,
during which the frequency of the QPO increased, the energy dependence
of the QPO became less depended on energy, and the soft phase lag
dropped from 2.5 radian to $\sim$1 radian (between the energy ranges
2.8--8.7 keV and 8.7--60 keV).  Unfortunately, this evolution occurred
during an Earth occultation of the source and a passage of the
satellite through the SAA, so a detail study of this process could not
be made.  From the X-ray colours and the HIDs and CD, it is clear that
a small but significant spectral difference is present between the two
parts of the observation, with the part containing the 5 Hz QPO
slightly harder than the part with the 3 Hz QPO.  This shows that a
very slight change of the X-ray spectrum is accompanied by a
significant change in the rapid X-ray variability. Whatever triggered
this change, it only minorly effected the X-ray spectrum. This
significant difference between the two parts of the observation also
makes it clear that during the VHS of GRS 1739--278, the accretion
processes involved are far from stable but they are very dynamic on a
time scale of less than an hour.

The difference between the first part and second part of the
observation was not reported by Borozdin \& Trudolyubov (2000), most
likely because they combined both parts together in their analysis
without performing any time selections. The 5 Hz QPO parameters quoted
by them are therefore contaminated by the inclusion of the first part
of the data which does not contain this feature. Therefore, we report
a significantly larger strength for the 5 Hz QPO and its first
overtone then Borozdin \& Trudolyubov (2000).

Recently, the phase lags of the low-frequency QPOs in BHC have
received a considerable amount of attention in the
literature. However, so far the phenomenology of these QPOs and their
phase lags in particular are far from understood. The lags have now
been studied for 1--20 Hz QPOs in GS 1124--683 (Takizawa et al. 1997),
XTE J1550--564 (Wijnands, Homan, \& van der Klis 1999; Cui et
al. 2000; Remillard et al. 2001), GRS 1915+105 (Reig et al. 2000; Lin
et al. 2000; Tomsick \& Kaaret 2001), XTE J1859+226 (Cui et al. 2000),
and GRS 1739--278 (this study).  The phase lags show a large variety
of behaviour. The different harmonics can have all the same sign for
the lags (e.g., GS 1124--683; GRS 1739--278; although the sign can be
both positive or negative) or can have different signs (the so-called
'alternating phase lags'; e.g., XTE J1550--564; GRS 1915+105). The
situation is made even more complex by the very complex evolution of
the phase lags in several sources (e.g., XTE J1550-564; GRS 1915+105).

Several theoretical studies have tried to address the complicated
phase lags behaviour of the low-frequency QPOs in BHCs (e.g., Nobili
et al. 2000; B\"ottcher \& Liang 2000\footnote{Although submitted to
ApJ for publication and available on astro-ph, this paper has been
retracted for publication due to two major problems with the proposed
QPO model: (1) the energy needed for the QPO mechanism to work
exceeded that of the available energy, and (2) the predicted energy
dependence of the QPO was the opposite to that observed (M. B\"ottcher
2001 private communication)}; B\"ottcher 2001), but those studies have
focussed on the QPOs in GRS 1915+105. It is unclear to what extent the
overall very complex behaviour of this source is influencing its QPO
behaviour. Extrapolation from models for the behaviour (i.e., the QPOs
and their phase lag behaviour) of GRS 1915+105 to other BHCs might
turn out to be difficult and subject to errors. At the moment, there
is no model available which can explain the observed phase lags of the
QPOs, the evolution of those lags for the individual sources, and the
differences between QPO behaviour in the different BHCs. However, it
is clear that a Comptonizing medium around the black-hole (thought to
produce the power-law tail in the spectrum) cannot explain the lags
observed for the QPOs around 6 Hz, either due to Compton up- or
down-scattering.

Recently, in several BHC transients QPOs above 100 Hz were found
(e.g., Remillard et al. 1999a,b; Cui et al. 2000; Homan et al. 2001;
Strohmayer 2001). Often, although not always, these QPOs are found
simultaneously with low-frequency QPOs ($<$10 Hz), with similar
characteristics as the 5 Hz QPO in GRS 1739--278. Therefore, similar
high-frequency QPOs might also be present in GRS 1739--278. However, a
search for such QPOs did not result in a significant detection, but
the upper limits on the presence of such QPOs are not very
stringent. Although in some sources stronger high-frequency QPOs have
been found (e.g., XTE J1859+226; Cui et al. 2000), in other sources
the high-frequency QPOs were considerable weaker than our upper limits
(e.g., XTE J1550--564; Homan et al. 2001). Therefore, we cannot
exclude that high frequency QPOs were present in GRS 1739--278.

\section*{Acknowledgments}

This work was supported by NASA through Chandra Postdoctoral
Fellowship grant number PF9-10010 awarded by CXC, which is operated by
SAO for NASA under contract NAS8-39073.  This research has made use of
data obtained through the HEASARC Online Service, provided by the
NASA/GSFC and quick-look results provided by the ASM/{\it RXTE} team.

\clearpage

\begin{table*}
\caption{QPO parameters versus photon energy$^a$ \label{tab:energy}}
\begin{flushleft}
\begin{tabular}{lllll}
\hline
\hline
Photon energy range & Fundamental amplitude & Overtone amplitude & Fundamental phase lag$^b$ \\
    (keV)            &          (\% rms)     &    (\% rms)        &
(rad)\\
\hline
\hline
   2--3.4           & $<$1.3                & $<$1.6             & -- \\
   3.4--4.0         & 1.67\pp0.18           & 1.59\pp0.29        & $-0.003$\pp0.02\\                
   4.3--5.0         & 2.68\pp0.23           & $<$1.2             & 0$^c$\\
   5.0--6.2         & 2.67\pp0.15           & $<$1.8             &  0.002\pp0.02\\                
   6.2--7.5         & 4.07\pp0.22           & 2.17\pp0.57        & $-0.03$\pp0.02\\
   7.5--9.3         & 4.42\pp0.28           & 4.10\pp0.43        & $-0.05$\pp0.02\\                
   9.3--11.2        & 5.01\pp0.56           & 6.27\pp0.69        & $-0.14$\pp0.03\\
  11.2--22.0        & $<$4.5                & 7.50\pp0.59        & $-0.22$\pp0.03\\
\hline
   3.4--5.6         & 2.10\pp0.08           & 1.32\pp0.18        & -- \\
   5.6--7.5         & 3.62\pp0.14           & 1.84\pp0.36        & -- \\
\hline
\hline
\multicolumn{5}{l}{$^a$ The errors on the fit parameters are for
$\Delta\chi^2 =1.0$ and the upper limits are for 95\% confidence levels}\\
\multicolumn{5}{l}{$^b$ Calculated for the 1.11 Hz interval (the FWHM of
the QPO) centred on the peak frequency of the QPO (5.09 Hz)}\\
\multicolumn{5}{l}{$^c$ Reference band}
\end{tabular}
\end{flushleft}
\end{table*}

\begin{table*}
\caption{The 3 Hz QPO vs. the 5 Hz QPO \label{tab:diff}}
\begin{flushleft}
\begin{tabular}{ccccc}
\hline
\hline
Frequency   & \multicolumn{3}{c}{Amplitude}                    & Phase lags$^a$  \\ \cline{2-4} 
            & 2--4.3 keV & 4.3--11.2 keV & 11.2--31.7 keV      &          \\
(Hz)        & (\% rms)   & (\% rms)      & (\% rms)            & (radian) \\
\hline
3.2\pp0.1   & $<$1.1     & $<$4.5        & 9.8$^{+0.8}_{-0.7}$ &  --2.4\pp0.2 \\
5.09\pp0.02 & 1.3\pp0.1  & 3.15\pp0.07   & $<$4.5              &  --1.01\pp0.06 \\
\hline
\hline
\multicolumn{5}{l}{$^a$ Between the energy bands 2.8--8.7 keV and 8.7--60 keV}
\end{tabular}
\end{flushleft}
\end{table*}


\begin{thebibliography}{}
              
\bibitem{}Borozdin, K. N., \& Trudolyubov, S. P.  2000, ApJ, 533, L131
              
\bibitem{}Borozdin, K., Alexandrovich, N., Sunyaev, R.  1996, IAUC,
6350
              
\bibitem{}Borozdin, K. N., Revnivtsev, M. G., Trudolyubov, S. P.,
Aleksandrovich, N. L., Sunyaev, R. A., Skinner, G. K.  1998, Astronomy
Letters, 24, 435 (astro-ph/9901192)
         

\bibitem{}B\"ottcher, M. 2001, ApJ, 553, 960     
              
\bibitem{}B\"ottcher, M. \& Liang, E. P. 2000, ApJ, submitted but
retracted (astro-ph/0003139)
              
\bibitem{}Cui, W., Shrader, C. R., Haswell, C. A., Hynes, R. I. 2000,
ApJ, 535, L123
              
\bibitem{}Dennerl, K., \& Greiner, J.  1996, IAUC, 6426

\bibitem{}Dieters, S. W., Belloni, T., Kuulkers, E., Woods, P., Cui,
W., Zhang, S. N., Chen, W., van der Klis, M., van Paradijs, J., Swank,
J., Lewin, W. H. G., Kouveliotou, C. 2000, ApJ, 538, 307
              
\bibitem{}Greiner, J., Dennerl, K., Predehl, P.  1996, A\&A, 314, L21
              

\bibitem{}Homan, J., Wijnands, R., van der Klis, M., Belloni, T., van
Paradijs, J., Klein-Wolt, M., Fender, R., M\'endez, M. 2001, ApJS 132,
377
              
\bibitem{}Hjellming, R. M., Rupen, M. P., Marti, J., Mirabel, F.,
Rodriguez, L. F.  1996, IAUC, 6384
              
              
\bibitem{}Leahy, D. A., Darbo, W., Elsner, R. F., Weisskopf, M. C.,
Kahn, S., Sutherland, P. G., Grindlay, J. E. 1983, ApJ, 266, 160

\bibitem{}Levine, A. M., Bradt, H., Cui, W., Jernigan, J. G., Morgan,
E. H., Remillard, R., Shirey, R. E., Smith, D. A. 1996, ApJ 469, L33

\bibitem{}Lin, D., Smith, I. A., Liang, E. P., B\"ottcher, M. 2000,
ApJ, 543, L141

              
\bibitem{}Marti, J., Mirabel, I. F., Duc, P.-A., Rodrigu\'ez, L. F.
1997, A\&A, 323, 158
              
\bibitem{}Mirabel, I. F., Marti, J., Duc, P.-A., Rodrigu\'ez, L. F.
1996, IAUC, 6428


\bibitem{}Nobili, L., Turolla, R., Zampieri, L., Belloni, T. 2000,
ApJ, 538, L137
              
              
\bibitem{}Paul, J., Bouchet, L., Churazov, E., Sunyaev, R.  1996,
IAUC, 6348

\bibitem{}Psaltis, D., Belloni, T., van der Klis, M. 1999, ApJ, 520, 262
              
              
\bibitem{}Remillard, R. A., McClintock, J. E., Sobczak, G. J., Bailyn,
C. D., Orosz, J. A., Morgan, E. H., \& Levine, A. M.  1999a, ApJ, 517,
L127
              
\bibitem{}Remillard, R. A., Morgan, E. H., McClintock, J. E., Bailyn,
C. D., Orosz, J. A. 1999b, ApJ, 522, 397


\bibitem{}Remillard, R. A., Sobczak, G. J., Muno, M. P., McClintock,
J. E. 2001, ApJ submitted (astro-ph/0105508)



\bibitem{}Reig, P., Belloni, T., van der Klis, M., M\'endez, M.,
Kylafis, N. D., Ford, E. C. 2000, ApJ, 541, 883

\bibitem{}Revnivtsev, M. G., Trudolyubov, S. P., \& Borozdin,
K. N. 2000, MNRAS, 312, 151
              
              
\bibitem{}Strohmayer, T.  2001, ApJ, 552, L49

\bibitem{}Takeshima, T., Cannizzo, J. K., Corbet, R., Marshall, F. E.
1996, IAUC, 6390

\bibitem{}Takizawa, M., et al. 1997, ApJ, 489, 272

\bibitem{}Tanaka. Y, \& Lewin, W. H. G.  1995, In: {\it X-ray
Binaries}, W. H. G. Lewin, J. van Paradijs, \& E. P. J. van den Heuvel
(eds.), Cambridge University Press, p. 126

\bibitem{}Tomsick, J. A. \& Kaaret, P. 2001, ApJ, 548, 401

\bibitem{}van der Klis, M.  1995, In: {\it X-ray Binaries},
W. H. G. Lewin, J. van Paradijs, \& E. P. J. van den Heuvel (eds.),
Cambridge University Press, p. 252
              
\bibitem{}van der Klis, M., Hasinger, G., Stella, L., Langmeier, A.,
van Paradijs, J., \& Lewin, W. H. G.  1987, ApJ, 319, l13
             

\bibitem{}Vargas M., et al. 1997, ApJ, 476, L23

\bibitem{}Wijnands, R., Homan, J., van der Klis, M. 1999, ApJ, 526, L33
              

\end{thebibliography}
\end{document}